\documentclass[twocolumn,aps,prx,superscriptaddress]{revtex4}
\usepackage{graphicx}
\usepackage{amssymb,amsmath}
\usepackage{bm}
\usepackage{dcolumn} 
\usepackage{subfigure}
\usepackage{float}
\usepackage[OT1]{fontenc} 
\usepackage{url}
\usepackage{slashed}
\usepackage{color}
\usepackage{verbatim}
\usepackage{times}
\usepackage{graphicx}
\usepackage[colorlinks,linkcolor=blue,anchorcolor=blue,citecolor=blue,hypertexnames=true]{hyperref}
\usepackage{url} 
\usepackage{multirow}

\begin{document}
	
\title{ Quantum state tomography with disentanglement algorithm}

\author{Juan Yao}
\email{juanyao.physics@gmail.com}
\affiliation{Shenzhen Institute for Quantum Science and Engineering, Southern University of Science and Technology, Shenzhen 518055, Guangdong, China}
\affiliation{International Quantum Academy, Shenzhen 518048, Guangdong, China}
\affiliation{Guangdong Provincial Key Laboratory of Quantum Science and Engineering, Southern University of Science and Technology, Shenzhen 518055, Guangdong, China}

\begin{abstract}
In this work, we report a novel quantum state reconstruction process based on the disentanglement algorithm. We propose a sequential disentanglement scheme, which can transform an unknown quantum state into a product of computational zero states. The inverse evolution of the zero states reconstructs the quantum state up to an overall phase.  By sequentially disentangling the qubits one by one, we reduce the required measurements with only single qubit measurement and identify the transformation unitary efficiently. Variational quantum circuit and reinforcement learning methods are used for the quantum circuit design for continuous and discrete quantum gates implementation. Demonstrations with our proposal for the reconstruction of the random states are presented. Our method is universal and imposes no specific ansatz or constraint on the quantum state. 
\end{abstract}
	
\maketitle

\section{Introduction}
The quantum state reconstruction through measurements on identical quantum states, known as quantum state tomography \cite{PhysRevLett.90.193601, PhysRevA.64.052312, MauroDAriano2003}, is crucial for benchmarking and verification of quantum system and quantum device. 
Although fully determination of the quantum state has already been achieved by using a complete set of observables
\cite{QuantumStateEstimation,Hou_2016}, 
the exponential scaling required in system size is inefficient. 
Various efficient learning techniques have been developed to address this issue, including machine learning methods utilizing neural networks 
\cite{science.aag2302,PhysRevA.102.022412,PhysRevLett.123.230504,PhysRevA.104.012401,Palmieri2020,Carrasquilla2019,Torlai2018},
Bayesian inference incorporating prior knowledge 
\cite{RevModPhys.81.299, PhysRevA.61.010304,PhysRevLett.108.070502, PhysRevA.95.062336, Granade_2016,Granade_2017,Blume-Kohout_2010},  
and tensor network representations
\cite{DBLP,YiKai2010,Lanyon2017,PhysRevLett.111.020401,PhysRevA.101.032321}.
One crucial feature that distinguishes quantum many-body systems from classical systems is the presence of entanglement, which arises from the correlations between different parts of the system \cite{RevModPhys.81.865}. The correlations among subsystems necessitate exponentially  large number of parameters in system size to fully describe the system state. Consequently, a quantum state absent of entanglement is significantly easier to characterize. In this context, we propose a novel protocol for quantum state tomography utilizing disentanglement algorithms. 

By transforming the unknown quantum state into a product state, the quantum state can be efficiently reconstructed by inverse transformation. In general, it is possible to evolve an unknown pure quantum state  $|\psi\rangle$ into a product state by an unitary transformation $\hat{\mathcal{U}}$. 
For an $N$-qubit quantum state to involve into a product state, it requires the removal of all the entanglement between any two subsystems. Ref.\cite{PhysRevApplied.13.024013} employs the fidelity between the whole final evolved state the product of zero states to quantify the disentanglement process.  Here we propose a disentanglement scheme in which the entanglement is sequentially removed. This sequential scheme offers two advantages: (i) only single-qubit measurements are required in each sequence, and (ii) the optimization process is much more efficient for each sequential process. 
During the disentanglement process, we need to identify the unitary transformation or the quantum circuit configuration. 
We present two methods to generate the unitary transformation. One relies on the variational quantum circuit, while the other utilizes the reinforcement learning method. In the following, we demonstrate the effectiveness of the sequential disentanglement scheme in efficiently reconstructing an unknown quantum state.

\section{ Method }
\begin{figure}[t]
\begin{centering}
\includegraphics[width=0.4\textwidth]{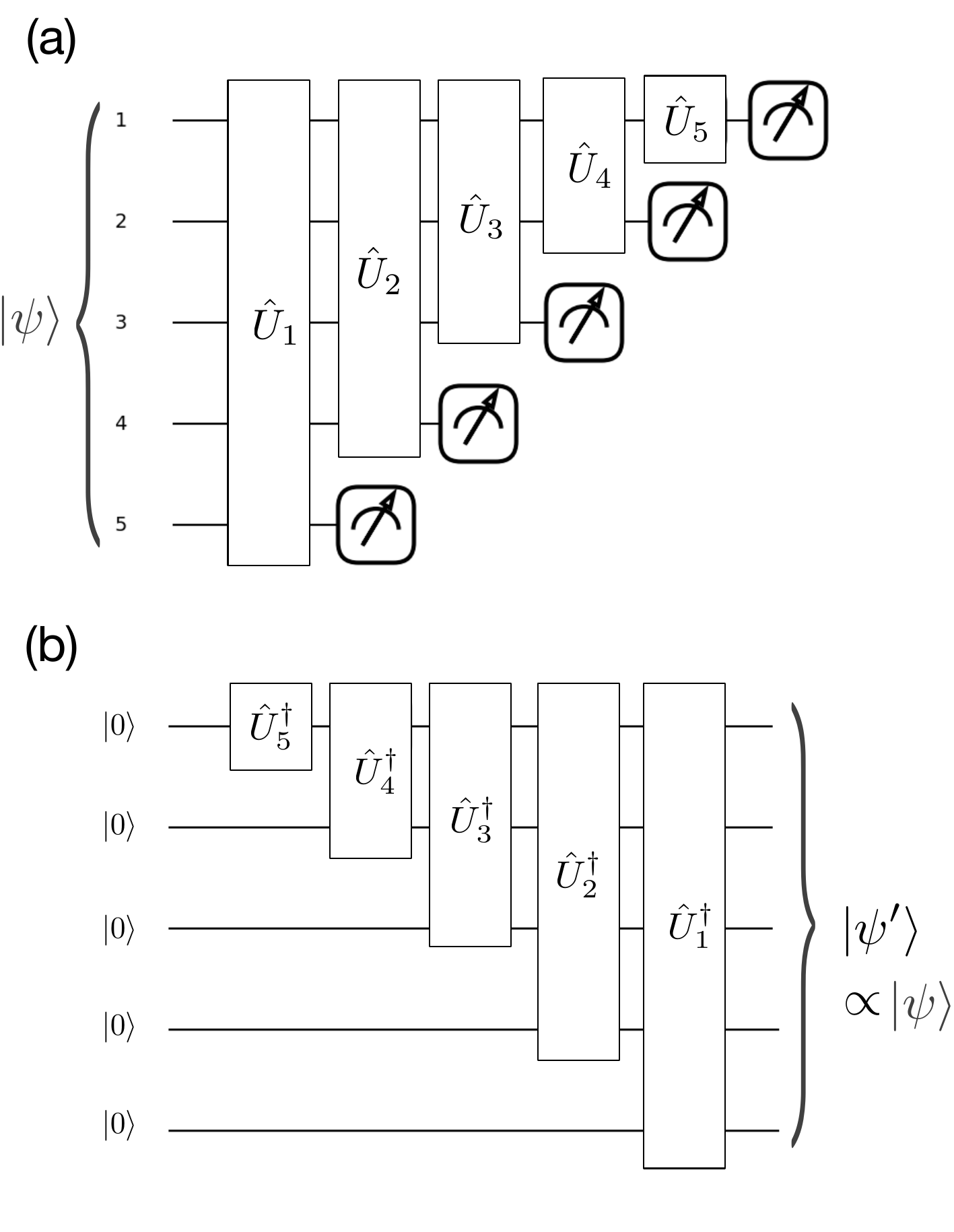}
\caption{(a) Sequential disentanglement scheme, demonstration with total number of qubits $N=5$. $|\psi\rangle$ represents the unknown quantum state to be reconstructed. The overall disentanglement transformation $\hat{\mathcal{U}}$ is composed by five disentangled circuits $\hat{U}_j$. (b) Reconstruction process from zero-product state, with the disentanglement transformation $\hat{\mathcal{U}}^\dagger$ identified with sequential disentanglement process in (a).  }
\label{Fig1}
\end{centering}
\end{figure}
Instead of directly identifying the whole unitary transformation $\hat{\mathcal{U}}$, we decompose $\hat{\mathcal{U}}$ into a sequence of  $\hat{U}_j$. Taking $N=5$ for example, as shown in Fig. \ref{Fig1} (a), we construct $\hat{U}_j$ such that 
\begin{equation}
|0\rangle^{\otimes N} \propto  ~  \hat{\mathcal{U}} |\psi\rangle = \Pi_{j=N}^{j=1} \hat{U}_{j} |\psi\rangle, 
\end{equation}
where $\propto$ represents a difference of an overall phase factor.  Two advantages arise from this sequential scheme: (i) only single-qubit measurements are required in each sequence;  and (ii) the optimization process is much more efficient for each sequential iteration. Within each sequence, the quantum circuit $\hat{U}_j$ is solely responsible for disentangling the last qubit of the entangled system, upon which the measurement is applied, as depicted in Fig. \ref{Fig1} (a),
\begin{equation}
m_q={\rm Tr}\{\hat{P}^q_0  \cdot  |\phi_j\rangle\langle\phi_j|\},
\end{equation}
where the corresponding projection operator $\hat{P}_0^q={\rm I}^{\otimes (q-1)}  \otimes  |0\rangle\langle0|$ and $m_q$ denotes the measured probability on the last qubit of the disentanglement system. Here $|\phi_j\rangle$ stands for the evolved state after disentanglement transformation $\hat{U}_j$. For each sequence, the loss function is defined as 
\begin{equation}
\mathcal{L} = 1- {\rm m}_q,
\label{lossEq}
\end{equation}
which aims to achieve $100\%$ probability of occupation of zero state for the measured qubit. Subsequently, the disentanglement transformation $\hat{U}_j$  can be determined based on this loss function using various optimizing strategies \cite{PhysRevA.101.052316, PhysRevApplied.13.024013, Schmalenpj2022, PhysRevLett.129.133601, PhysRevA.105.032427}.
Once the optimized $\hat{U}_j$ is identified in current sequence with infinitesimally small $\mathcal{L}$,   the evolved state can be casted as 
\begin{equation}
|\phi_j\rangle= |\psi_{j+1}\rangle \otimes |0\rangle.
\end{equation}
A $100\%$ probability of occupation with $\mathcal{L}=0$ guarantees the disentanglement between the subsystem and the last qubit. Thus in the next sequence, the disentangled system can be  given by 
\begin{equation}
\hat{\rho}_{\rm entangled}=|\psi_{j+1}\rangle\langle\psi_{j+1}|= {\rm Tr}_q\{ |\phi_j\rangle\langle \phi_j|\}  , 
\label{rho}
\end{equation}
where the partial trace is applied to the measured qubit.  Each disentanglement process generates the associated transformation $\hat{U}_j$. 

Overall, identifying the disentanglement transformation $\hat{U}_j$ is much easier compared with identifying $\hat{\mathcal{U}}$. In comparison to optimizing all parameters within $\hat{\mathcal{U}}$ together, the number of parameters in each sequence $\hat{U}_j$ is significantly smaller.  Moreover, within each sequence, the optimization target is only one qubit measurement result, which is much easier to achieve compared to targeting all the qubit measurement results.  Optimization process in each sequence is more focused and efficient. Thus, through decomposing $\hat{\mathcal{U}}$ into $\hat{U}_j$,  there is a significant reduction in the difficulties of removing entanglement. More specifically, when the disentangled system size is large, for instance $j=1$,  redundancy due to free choices of other qubit states will benefit the optimization process. When the disentangled system is small, for instance $j=5$, although benefits from redundancy decay, in this case, the small number of optimized parameters will facilitate the optimization process {(see appendix C for more details)}. 
Finally, a complete sequential disentanglement scheme, shown in Fig. \ref{Fig1} (a), will transform the quantum state $|\psi\rangle$ into a product state $|0\rangle^{\otimes N}$ using identified $\hat{U}_j (j=1,2,...,N)$. 

Noting that the overall phase factor cannot be resolved by the measurement, the unknown quantum state $|\psi\rangle$ can still be reconstructed by product of zero states $|0\rangle^{\otimes N}$. 
As shown in Fig. \ref{Fig1} (b), we can reconstruct the quantum state $|\psi^\prime\rangle$ by back propagation using the identified unitary transformations as 
\begin{equation}
|\psi^\prime\rangle = \Pi_{j=1}^{j=N}\hat{U}_i^\dagger |0\rangle^{\otimes N}. 
\end{equation}
The reconstructed quantum state $|\psi^\prime\rangle$ differs from the quantum state $|\psi\rangle$ up to an overall phase factor, which is irrelevant for physical observables measurements.  As the key point here is to identify the sequence of $\hat{U}_j$,  below, we adopt two strategies to construct the configuration of $\hat{U}_j$. The first one relies on the variational quantum circuit for the availability of continuous parametric single-qubit gate. The second one relies on reinforcement learning for limited discrete quantum gates. 

\section{Result}
\subsection{Variational quantum circuit}

\begin{figure}[h!!]
\begin{centering}
\includegraphics[width=0.35\textwidth]{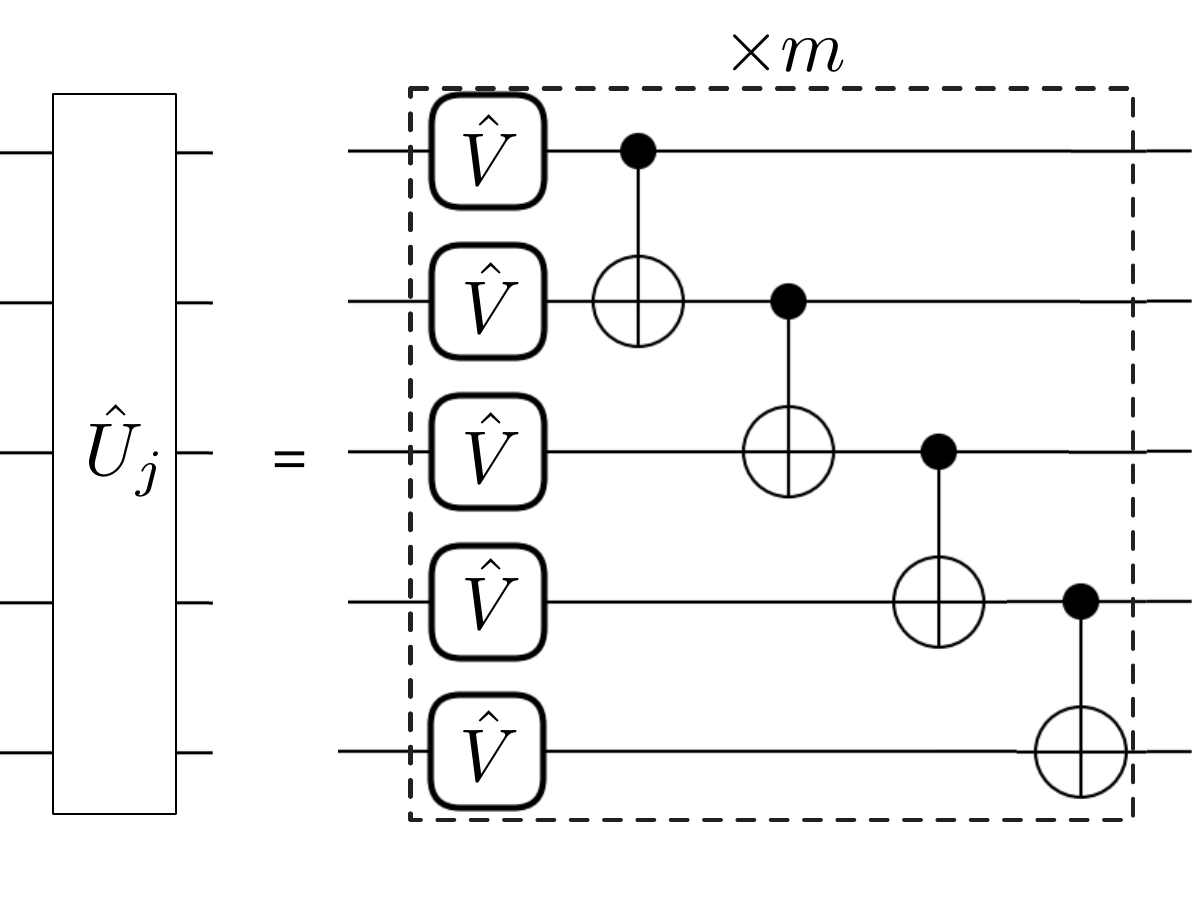}
\caption{Quantum circuit configuration for the parametric quantum circuit $\hat{U}_{j=1}$ with $N=5$ qubits.  The building block (dashed box) is composed by product of $N=5$ parametrized single-qubit unitary rotations $\hat{V}$ and $N-1$ CNOT gates. Multiple copies of these building blocks compose the variational quantum circuit $\hat{U}_1$. }
\label{Fig2}
\end{centering}
\end{figure}
In this section, we assume the availability of parametrized single-qubit gates, and two-qubit CNOT gates. To achieve disentanglement, we construct a quantum circuit for each sequence following the structure outlined in Fig. \ref{Fig2}.  For an $N$-qubit entangled system, the building block consists of $N$ single-qubit gates and $N-1$ CNOT gates. Using multiple copies of the building blocks enhances the disentanglement ability, which provides the basic configuration of the variational quantum circuit. The training parameters in the single-qubit gate are explicitly given by  
\begin{equation}
\hat{V}(\phi,\theta,\omega) = \begin{bmatrix}
\cos\frac{\theta}{2}e^{-i\frac{\phi+\omega}{2}}  &  -\sin\frac{\theta}{2}e^{i\frac{\phi-\omega}{2}}  \\
\sin\frac{\theta}{2}  e^{-i\frac{\phi-\omega}{2}} & \cos\frac{\theta}{2}e^{i\frac{\phi+\omega}{2}}  
\end{bmatrix}.
\end{equation}

Taking $N=5$ quantum state for demonstration, the training results are displayed in Fig. \ref{Fig3}.  Initially, we randomly generate an unknown quantum state $|\psi\rangle$ {(see appendix B for more details)}.  We simulate the optimization process using gradient descent with the PyTorch Adam optimizer.  Red lines with open circles display the loss function $\mathcal{L}$ during each training sequence. 
Due to the initial entanglement in the quantum state, $\mathcal{L}$ starts with a finite value at the beginning of training.
After training for multiple epochs, the loss function tends towards infinitesimally small values.
As $\mathcal{L}$ approaches zero, the measurement result for the last qubit consistently yields the computational zero state, indicating successful disentanglement. Completion of the training sequence is indicated by the minimization of $\mathcal{L}$ for $j-th$ qubit, as shown in Fig. \ref{Fig3}.  In the next sequence, similar training procedures are then applied to $\hat{U}_{j+1}$, providing the fixed quantum circuit $\hat{U}_j$ identified in the previous sequence. By repetition these disentanglement procedures, we gradually identify the quantum circuits $\hat{U}_j$ one by one.

To monitor the disentanglement process, in addition to $\mathcal{L}$, the purity of the system is calculated during training as 
\begin{equation}
\mathcal{P}= {\rm Tr} \{  \hat{\rho}_{\rm sub}^2 \}.
\label{EqPurity}
\end{equation}
Here, $\hat{\rho}_{\rm sub}={\rm Tr}_q|\phi_q\rangle\langle\phi_q|$, where the last qubit is partially traced out. 
In each training sequence, the corresponding purity $\mathcal{P}$ of the reduced subsystem is plotted  in Fig. \ref{Fig3} with solid line. When there is no coupling between the last qubit and the subsystem, $\mathcal{P}$ approaches one, indicating that both are pure states. Now, $\hat{\rho}_{\rm sub}$ serves as the entangled quantum system $\rho_{\rm entangled}$ for the next optimization sequence.  At the end of training for all sequences, both the loss functions $\mathcal{L}$ and purities converge.  The training results indicate the removal of all the correlations of the system. Due to the unresolved phase factor for each single qubit,  the overall evolved state will be the product state of single qubit $|0\rangle$ with a phase factor. 

\begin{figure}[h!!]
\begin{centering}
\includegraphics[width=0.45\textwidth]{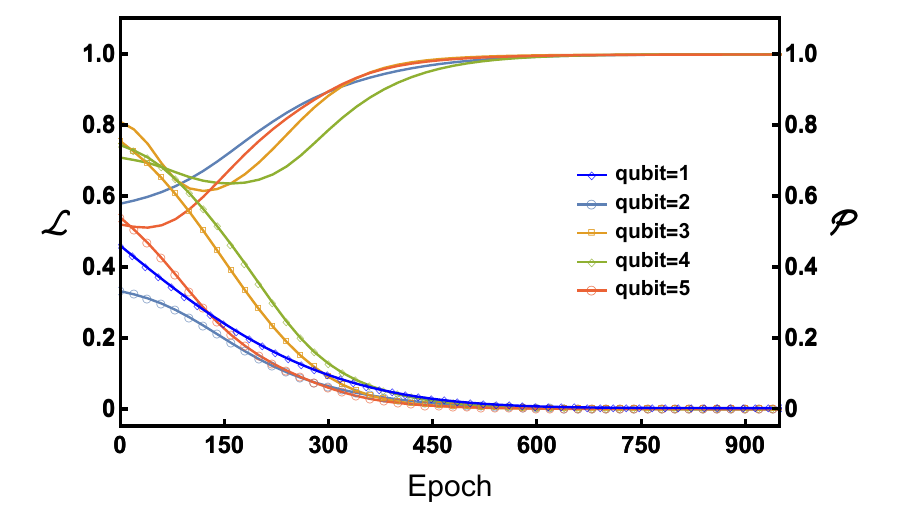}
\caption{ Training loss for all sequences with a quantum state having $N=5$ qubits. For the first sequence, with the disentangling qubit $j=5$, the red line with circles represents the trajectory  of the loss $\mathcal{L}$, while the red line $\mathcal{P}$ denotes the purity of the subsystem. Similar results are presented for other sequences.}
\label{Fig3}
\end{centering}
\end{figure}

\begin{figure}[h!!]
\begin{centering}
\includegraphics[width=0.43\textwidth]{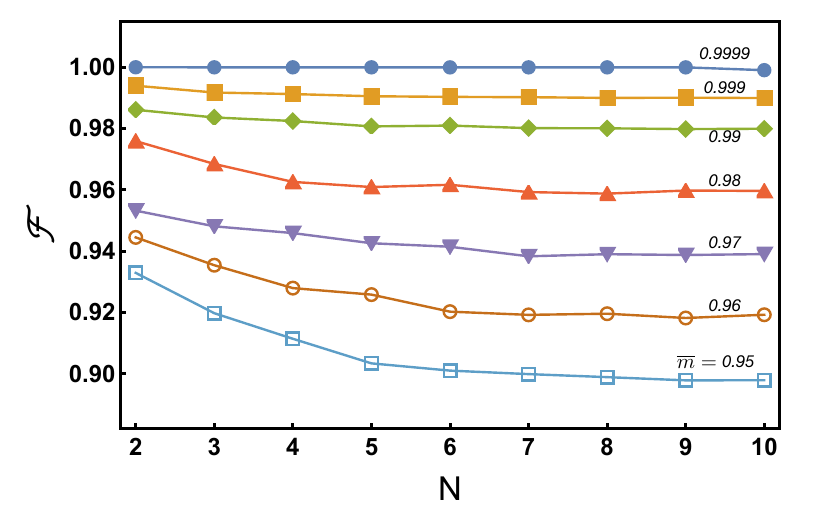}
\caption{At various training precision $\overline{m}$, plotting the reconstruction fidelity $\mathcal{F}$ as a function of system size $N$. } 
\label{Fig4}
\end{centering}
\end{figure}
Now, the initial unknown quantum state $|\psi\rangle$ can  be reconstructed using product of zero states $|0\rangle^{\otimes N}$ through back propagation, as illustrated in Fig. \ref{Fig1} (b).  
With the identified unitary transformations, the reconstructed quantum states $|\psi^\prime\rangle$ can be calculated as 
\begin{equation}
|\psi^\prime\rangle = \Pi_{j=1}^{j=N}\hat{U}_i^\dagger |0\rangle^{\otimes N}. 
\end{equation}
We plot the fidelity  $\mathcal{F}= |\langle \psi |\psi^\prime\rangle|^2$ between the reconstructed state and the unknown quantum state as in Fig. \ref{Fig4}. 
For real experimental platforms, during the optimization process, the loss function $\mathcal{L}$ cannot be exactly zero {due to measurement imperfections which may caused by noise or decoherence effects.} It is instructive to analyze the stability of the reconstruction fidelity with measurement errors.   As shown in Fig. \ref{Fig4}, the fidelity is plotted against different system size $N$ for various levels of training precisions. Let $\overline{m}=1/N\sum_q {\rm m}_q$ denote the average measurement outcome for all sequences. 
{We assume a positive correlation between the training error and the measurement imperfections. }
Setting a high training precision level with $\overline{m}=0.9999$, $\mathcal{F}$ is approximately $1$  for all system size $N$. In this case, we can reconstruct the unknown quantum state exactly in principle.  When the noise effects manifest with a low training precision level, as presented in Fig. \ref{Fig4}, the fidelity deviates from one with smaller $\overline{m}$.  Overall, the deviation is stronger for larger system size $N$. To reach a specific fidelity,  higher precision is required for larger system. Fidelities presented in Fig.~\ref{Fig4} provide a global picture of the stability of the reconstruction.

\subsection{Reinforcement learning for discrete gates}
For experimental implementation on quantum platforms with limited gate operations, parametrization may not be available in this case. Therefore, optimization of previously implemented variational quantum circuit  cannot be applied to this discrete problem. We will utilize the reinforcement learning strategy \cite{712192} for the quantum circuit design, given limited discrete gate operations (see appendix \textcolor{red}{A} for more details). Reinforcement learning, as a machine learning paradigm, can train a policy to make a series of decisions in an environment in order to maximize a reward signal \cite{NIPS1999_464d828b}. 
\begin{figure}[ht!!!]
\begin{centering}
\includegraphics[width=0.5\textwidth]{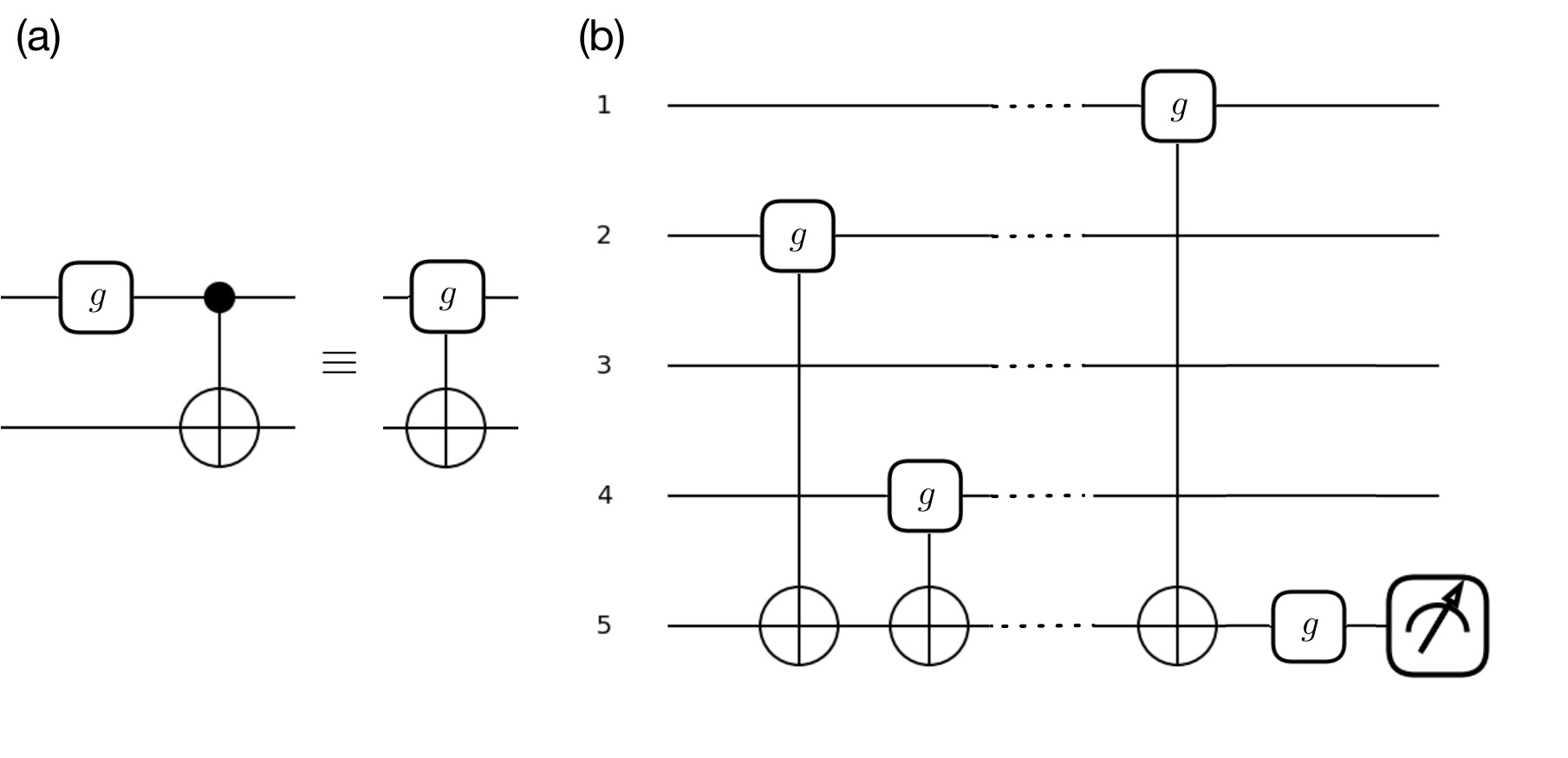}
\caption{(a) The action $a_j$ consists of a single-qubit gate $g\in \mathcal{G}$ followed by a CNOT gate. Explicitly, the operation can be given by  ${\rm CNOT} \cdot g \otimes {\rm  I} $. (b) An illustration of the quantum circuit configuration generated by the reinforcement learning method. In each layer, one action $a_j$ is sampled from the action set $\mathcal{A}$.  }
\label{Fig5}
\end{centering}
\end{figure}
Given a random entangled quantum state, the trained policy is likely to generate the configuration composing the disentanglement circuit $\hat{U}_j$ with high probability. 

We assume the following available single-qubit quantum gates:
\begin{equation}
\mathcal{G} = \{ {\rm I , X, Y, Z, H, T, S} \}, 
\label{EqG}
\end{equation}
and the two-qubit CNOT gate. Here ${\rm I}$ stands for identity,  ${\rm X, Y, Z}$ are three Pauli gates, H is the Hadamard gate and S is the phase gate with ${\rm T}^2=S$.  Within the framework of reinforcement learning, for the sake of simplicity in training,  we formulate the action as combined two-qubit gates. As illustrated in Fig. \ref{Fig5} (a), each  action consists of a single-qubit gate followed by a CNOT gate. The action set consists of seven different combined two-qubit gates. 
For each two-qubit gate,  we impose restrictions such that the first qubit operation can be applied to any qubit except the measurement qubit, and the second qubit operation can only be applied to the disentangling qubit to be measured.  
Thus, for disentangling an $N$ qubit quantum state, there are $N$ ways of applying qubit operation. Taking into account of seven types of gates, the total number of possible actions is $d=7N$. Then the action set can be denoted as $\mathcal{A}=\{a_j|j=1,2,...,d\}$.   

In the reinforcement learning algorithm, we use the classical neural network as the policy function, which maps the states to the actions as shown in Fig. \ref{FigApp1}. For each input state, the classical neural network generates the probability distribution $P(a_s)$ for all the possible actions. The next action is sampled according to the probability distribution. The purpose of the constructed quantum circuit is to disentangle the last qubit of the system. It is reasonable to set  the measured probability after applying $a_s$ as the the reward function. 
To maximize the reward or improve the measurement probability, the loss of the policy function is given by 
\begin{equation}
loss =  - \sum_s \log[ P(a_s)] \mathcal{R}(a_s),
\label{Eqloss}
\end{equation}
where $\mathcal{R}(a_s)$ is the reward function of action $a_s$. Training policy with the above loss function results in higher probability of action $a_s$ for actions with larger reward $\mathcal{R}(a_s)$. This means that actions with larger rewards will have higher probabilities after training. 
The reinforcement learning procedures are as follows: 
(i) Training dataset generating: generating $\mathcal{D}$ action series with a length of $L$ according to the policy, and calculating the corresponding rewards for each action; 
(ii) Improving the policy performance:  training the policy (classical neural network) using the dataset generated in (i) with loss given by Eq. \eqref{Eqloss}. 
Repeating procedure (i) and (ii) until the reward of the action reaches the maximal value $100\%$, resulting in the disentanglement of the corresponding qubit from the system. In this circumstance, we can disentangle a quantum state sequentially with the above reinforcement learning procedures. 

\begin{figure}[t]
\begin{centering}
\includegraphics[width=0.45\textwidth]{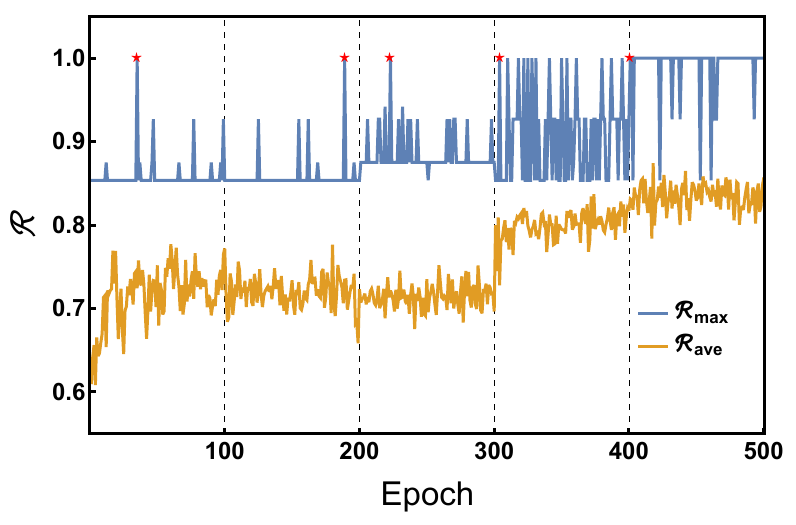}
\caption{ Maximal reward (blue line) and average reward (yellow line) calculated during the reinforcement learning procedure. Every $100$ training epochs, one qubit will be disentangled from the system. For $N=5$ entangled quantum state, each red star point provides the action configuration for the disentanglement quantum circuit design. }
\label{Fig6}
\end{centering}
\end{figure}
As illustrated by Fig. \ref{Fig5} (b), we set the length of disentanglement circuit to $L=10$, the dataset dimension to $\mathcal{D}=50$, and the training epoch to $T=100$. Reinforcement learning procedures are applied to each qubit sequentially for a random generated $N=5$ entangled quantum state. Each qubit will be disentangled within $T=100$ training epochs. As shown in Fig. \ref{Fig6}, we plot the rewards of the actions in the dataset generated in step (i) during training. Focusing on the first $T=100$ training epochs (the first disentanglement sequence),  the average rewards (yellow line) increase and the the maximum reward $\mathcal{R}_{\rm max}(a_s)=1$ is achieved. 
Similar results are presented for other disentanglement sequences in Fig. \ref{Fig6}. 
It should be noted that multiple instances of $\mathcal{R}_{\rm max}=1$ may appear during training. 
Multiple configurations exist for constructing the disentanglement transformation. In practice, we can stop the training process when the first maximum reward $\mathcal{R}=1$ is obtained, as denoted by the red star points in Fig. \ref{Fig6}. 
The five disentanglement action series, denoted by the red stars, can be used to reconstruct the input unknown quantum state.  It is worth noting that successfully identifying a disentanglement quantum neural network configuration is not always guaranteed, especially when the action length $L$ is too short or when the available  gate types are too limited to implement a disentanglement unitary transformation.

\section{Summary and Outlook} 
We propose a novel quantum state reconstruction process based on the disentanglement algorithm. By using the single-qubit measurement result, we can identify the disentanglement transformation. We identify the transformation using variational quantum circuit optimization and reinforcement learning-assisted circuit design. 
Both methods can successfully disentangle and reconstruct the target quantum state, as well as implement the quantum state tomography task.  
{The introduced sequential disentanglement scheme only requires single-qubit measurements and can be optimized efficiently, eliminating the need for complex post-processing procedures.} Our method is universal and imposes no specific ansatz or constrains on the quantum state, making it applicable to various circumstances. 

{In the main text, we assume the availability of variational quantum gates.  Although we primarily assume the use of single-qubit parameterized gates, a two-qubit gate is also required to enable the disentanglement capabilities, which is not restricted to CNOT. Any two-qubit gate that can provide entanglement can be adopted. If parameterized two-qubit gates are available, more compact quantum circuit configurations can also be designed. }

\vspace{5pt}
 \emph{Acknowledgement.} 
 We thank Yadong Wu, Pengfei Zhang, Xiuhao Deng, Georg Engelhardt {and Tao Xin} for helpful discussions. 
This work is supported by the Science, Technology and Innovation Commission of Shenzhen, Municipality (KQTD20210811090049034), Guangdong Basic and Applied Basic Research Foundation (2022B1515120021), and  
National Natural Science Foundation of China (Grant No. 11904190).

\appendix
\section{Reinforcement learning}
In this section, we explain the basic formalism of the reinforcement learning (RL) method adopted in the main manuscript.
Reinforcement learning, as a machine learning paradigm, trains a policy to make a series of decisions in an environment with the goal of maximizing a reward signal \cite{NIPS1999_464d828b}. The environment is characterized by a state space $\mathcal{S}$, while all the possible decisions constitute the action space $\mathcal{A}$. The reward $\mathcal{R}$ evaluates the score of action $a_j$ generated according to the policy.

\begin{figure}[ht]
\begin{centering}
\includegraphics[width=0.4\textwidth]{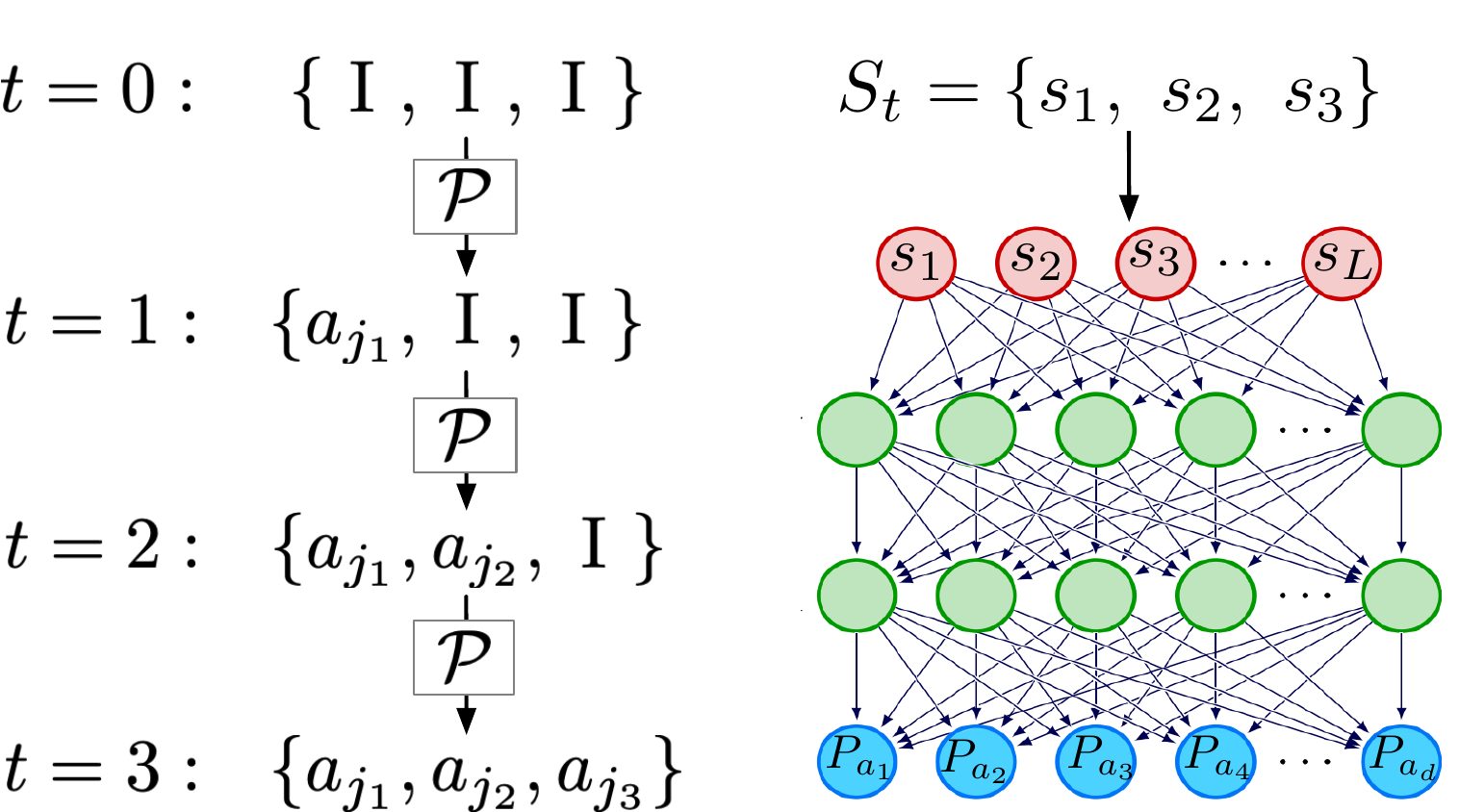}
\caption{ A configuration of disentanglement circuit generated by the policy.  The policy is implemented by a classical neural network.}
\label{FigApp1}
\end{centering}
\end{figure}

The goal of each disentanglement sequence is to identify a series of gate operations that disentangle the last qubit.
As shown in Fig. \ref{FigApp1}, we outline the process of generating the disentanglement configuration.  In the context of reinforcement learning (RL), the quantum state is represented by a series of gate operations with length $L$.
Each set $S_t={s_1,s_2,...s_L}$ represents a quantum state $|\psi_t\rangle$, where $|\psi_t\rangle = \Pi_{l=1}^{L}s_l|\psi\rangle $.
Initially,  the quantum state $|\psi\rangle$ is presented by the set $S_0=\{ {\rm I},{\rm I},...,{\rm I}\}$. 
As illustrated in Fig. \ref{FigApp1}, we consider a disentanglement circuit with a layer length of $L=3$. At the initial time step $t=0$, we start with the initial state $S_0$ and generate the first gate operation $a_{j_1}$ using policy $P(a_j)$.
Similarly, at $t=L$, all gate actions are generated according to the policy. The set $S_L$ represents a disentangled configuration.

In the following, we will introduce how to train the policy using reward functions. As illustrated in Fig. \ref{FigApp1}, the policy is implemented by a classical neural network. 
The input of the neural network is the state $S_t$, and its output is the probability distribution over all possible actions. Typically, actions are sampled from this distribution according to $P(a_j)$. After each action applied to the environment, we use rewards to evaluate action's performance. The goal of the training is to maximize the reward. During training, the loss function is defined as the weighted sum of the negative logarithm of the action probabilities multiplied by their corresponding rewards:
\begin{equation}
loss =  - \sum_s \log[ P(a_s)] \mathcal{R}(a_s),
\end{equation}
where $s$ iterates over multiple configurations of the disentanglement circuit.
If the size of action space is $d$ and the length of the circuit is $L$, the total number of possible states  is $d^L$.  

\section{Random quantum state generation}
In the main text, a random quantum state is generated for demonstration of the sequential disentanglement protocol. There are multiple methods to generating a quantum state. Here, we directly sample the coefficients of each computational basis $|j\rangle$ as follows: 
\begin{equation}
|\psi\rangle = \sum_j \tilde{\alpha}_j |j \rangle,
\end{equation}
where coefficients $\tilde{\alpha}_j$ are generally complex numbers. Each coefficient can be rewritten as the sum of its real and imaginary parts separately,  $\alpha_j = a_j+i b_j$. Numerically, $a_j$ and $b_j$ are sampled randomly between $0$ and $1$.  In most cases, a set of coefficients $\{ \alpha_j|j=1,...,2^N\}$ dose not normalize to 1. Therefore, a normalization process is needed to obtain the quantum state coefficients set $\{ \tilde{\alpha}_j= \alpha_j/\sqrt{\mathcal{N}}|j=1,...,2^N\}$ with normalization factor $\mathcal{N}= \sum_j \alpha_j\alpha_j^*$. 

\section{Training efficiency comparison}
\begin{figure}[ht]
\begin{centering}
\includegraphics[width=0.4\textwidth]{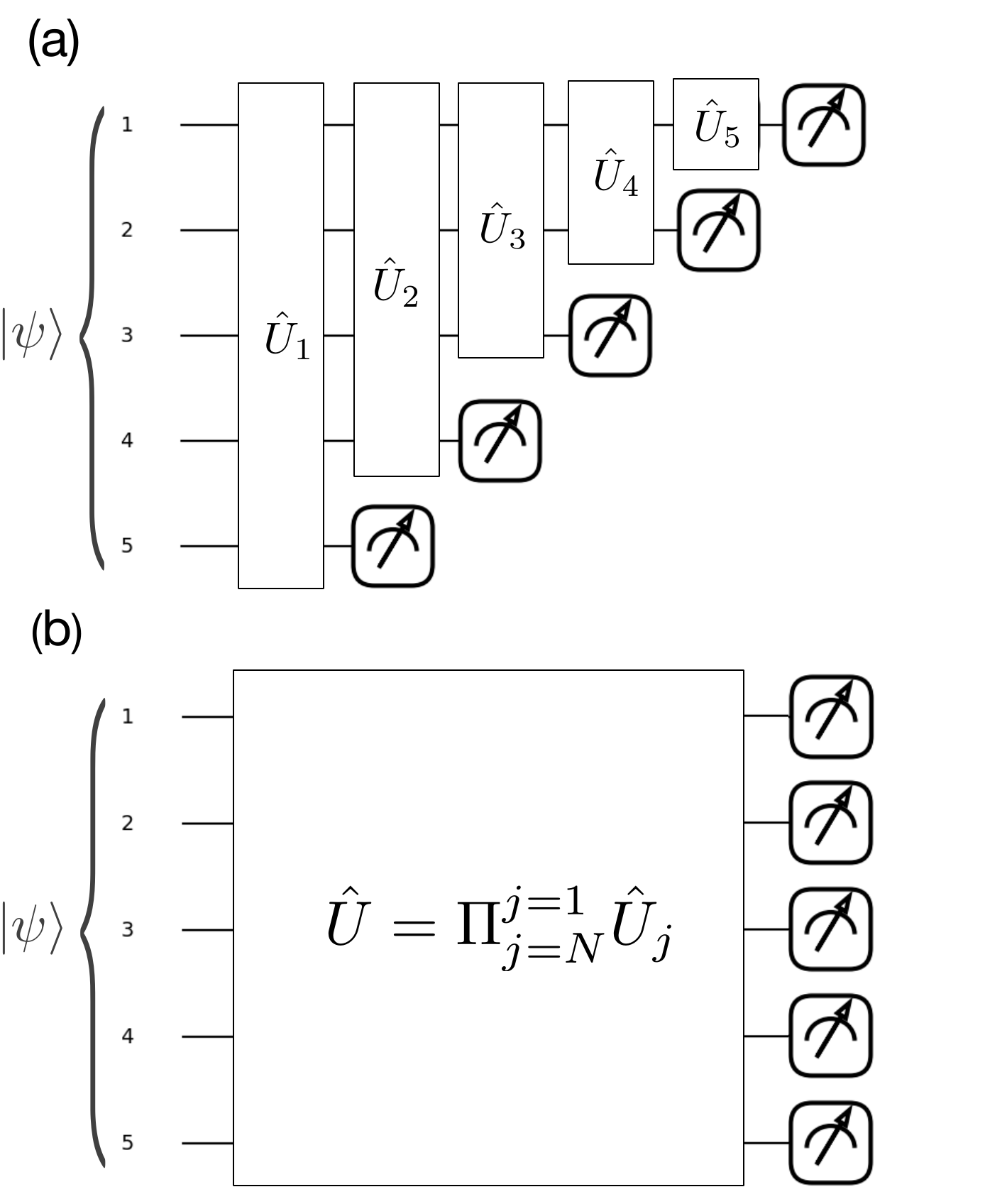}
\caption{ Two different disentanglement schemes for $N=5$ for demonstration.  (a) A full final measurement is applied in the end of quantum circuit. Updating the parameters with probabilities for all qubit measurement. (b) A single qubit measurement is applied within each sequence for disentangling a specific single qubit. }
\label{FigApp2}
\end{centering}
\end{figure}

As shown in Fig. \ref{FigApp2}, we employ two different disentanglement schemes to demonstrate the training efficiency of our sequential protocol. 

In Fig. \ref{FigApp2}(b), each $U_j$  is constructed using multiple copies of the building block provided in the main text, as illustrated in Fig. \ref{Fig2}. The number of copies, $m$, is set as:
\begin{equation}
m = N_s \times r
\end{equation}
where $N_s$ represents the subsystem size and $r$ is the manually set repetition factor. By increasing $r$, we extend the depth of the quantum circuit, thereby increasing the number of associated variational parameters.
In contrast, Fig. \ref{FigApp2}(a) adopts a non-sequential scheme, where a final full measurement is applied for gradient descent by requiring all measurement probabilities to equal one. For comparison purposes, we configure the quantum circuit in both schemes to be the same, with $\hat{U}= \Pi_{j=N}^{j=1}\hat{U}_j$.

In the sequential scheme, the parameters within each $\hat{U}_j$ are optimized using a loss function based on single-qubit measurements, as described by Eq. \ref{lossEq}. 
For instance, considering $N=8$ and $r=5$, and setting the training loss to $10^{-4}$, the required training epochs for optimization for each sequence are summarized in Table \ref{table1}. 
\begin{table}[ht!]
\centering
\begin{tabular}{ |c||c|c|c|c|  }
 \hline
 \multicolumn{5}{|c|}{$N=8$ and  $r=5$} \\
 \hline 
 ~~~~j~~~~& ~~~~gates~~~~ &  ~parameters~ & ~~epochs~~ & ~~~~~~~$S_{GD}$~~~~~~~ \\
 \hline
1  & 320    &960&   332& 318,720 \\
2&   245  & 735   &239 & 175,665 \\
3 &180 & 540&  205 & 110,700\\
4    &125 & 375&  181 & 67,875\\
5&   80  & 240&189 & 45,360\\
6& 45  & 135  &149 & 20,115\\
7& 20  & 60   &398 & 23,880\\
8& 5  & 15   &815 & 12,225 \\
 \hline
 ~~Total~~ & 1,020  &3,060 &  &~774,540~\\
 \hline
\end{tabular}
\caption{Collecting statistical training data includes tracking the number of parameters, the number of training epochs, and the number of gradient descent (GD) steps. }
\label{table1}
\end{table}
In the first column of Table \ref{table1}, $j$ indexes the sequence number. The second column provides the number of single-qubit gates. The third column lists the corresponding number of parameters to be optimized. The fourth column shows the cumulative training iterations when the loss function reaches $10^{-4}$. The final column indicates the required number of gradient descent computations, which is the product of the number of parameters and the number of epochs in each row. 
\begin{figure}[th]
\begin{centering}
\includegraphics[width=0.45\textwidth]{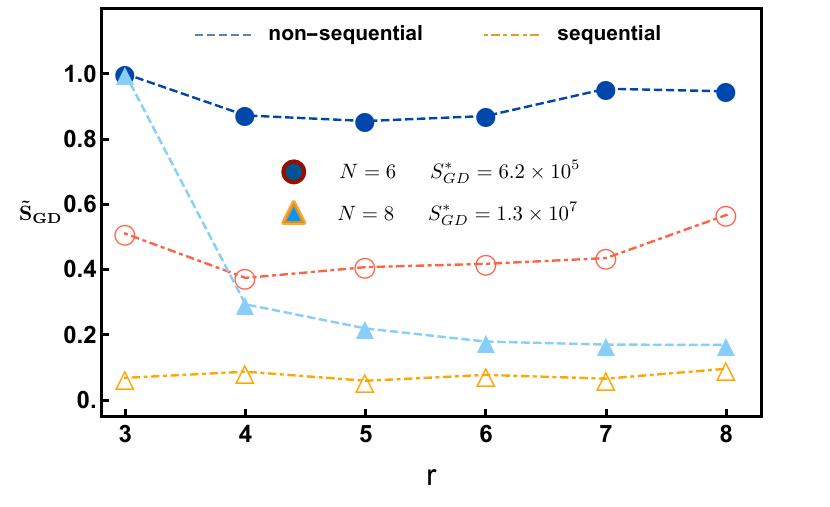}
\caption{ Number of gradient decent steps calculated in the training process. $\tilde{S}_{GD} $ is normalized by the maximal $S_{GD}$ for non-sequential scheme.  When $N=6$, $S_{GD}^{*} = 6.2\times 10^5$. When $N=8$, the maximal $S_{GD}$ is $1.3\times10^7$. }
\label{FigApp3}
\end{centering}
\end{figure}

In comparison, the non-sequential scheme trains all the parameters in $\hat{U}$ simultaneously using a loss function defined by the summation of all the measurement probabilities. This approach results in $S_{GD}$ reaching as high as 2,867,220, which is 2.7 times more than that of the sequential scheme. Furthermore, as shown in Fig. \ref{FigApp3}, we plot the normalized $\tilde{S}_{GD}$ in terms of the repetition times $r$ for $N=6$ and $N=8$. In both cases, the sequential scheme demonstrates significantly improved training efficiency.

\bibliographystyle{unsrt}  

\end{document}